# Car-Sharing Subscription Preferences: The Case of Copenhagen, Munich, and Tel Aviv-Yafo

**Authors:**

Monteiro, Mayara Moraes[1]; Azevedo, Carlos M. Lima[1]; Kamargianni, Maria[2]; Shiftan, Yoram[3]; Gal-Tzur, Ayelet[3]; Tavory, Sharon Shoshany[3]; Antoniou, Constantinos[4]; Cantelmo, Guido[1].

[1]*Technicala University of Denmark,* [2]*University College London,* [3]*Technion – Israel institute of Technology,* [4]*Technicala University of Munich.*

**Abstract**

Car-sharing services have been providing short-term car access to their users, contributing to sustainable urban mobility and generating positive societal and often environmental impacts. As car-sharing business models vary, it is important to understand what features drive the attraction and retention of its members in different contexts. For that, it is essential to examine individuals' preferences for subscriptions to different business models and what they perceive as most relevant, as well as understand what could be attractive incentives. This study aims precisely to examine individuals' preferences for the subscription of different car-sharing services in different cities. We designed a stated preference experiment and collected data from three different urban car-sharing settings, namely Copenhagen, Munich, and Tel Aviv-Yafo. Then a mixed logit model was estimated to uncover car-sharing plan subscription and incentives preferences. The results improve our understanding of how both the features of the car-sharing business model and the provision of incentives can maintain and attract members to the system. The achieved insights pave the road for the actual design of car-sharing business models and incentives that can be offered by existing and future car-sharing companies in the studied or similar cities.

**1 Introduction**

Accumulating more than 32 million members, distributed across 47 countries and six continents (2018), car-sharing services have been providing short-term car access to their users (Shaheen et al., 2020). Under the umbrella of sharing economy solutions, car-sharing aims at encouraging sustainable urban mobility, shifting the focus from personal ownership to demand-fulfillment shared use (Mi and Coffman, 2019). Positive societal and environmental impacts derived from car-sharing solutions in cities include an increase in mobility flexibility (Clewlow, 2016) and the use of some alternative transportation modes (E. Martin and Shaheen, 2011), as well as reductions in car ownership (fewer resources required for mobility) (Clewlow, 2016)(Giesel and Nobis, 2016), kilometers traveled (Clewlow, 2016)(E. W. Martin and Shaheen, 2011), greenhouse gases and air pollutants emissions (E. W. Martin and Shaheen, 2011)(Chen and Kockelman, 2016), congestion (Alisoltani et al., 2021) and parking demand (Millard-Bal et al., 2005).

Today, car-sharing business models vary, among others, according to the level of flexibility on pick-up and return locations (e.g., station-based or free-floating services), ownership of the service/cars (e.g., private, cooperative, peer-to-peer), the composition of car fleet (e.g., electric, combustion, hybrid cars; utilitarian, luxury, small city cars), pricing scheme (e.g., minutes, hourly and daily packages) and parking opportunities (e.g., public or private reserved parking spaces) (Shaheen et al., 2019). The appropriate combination of these features can attract and retain users/members, which are essential for the presence of car-sharing services in cities.

Yet, the analysis of how car-sharing service features affect car-sharing membership is limited. To examine the determinants of car-sharing subscription, previous studies have mainly focused on how socio-demographic characteristics influence the likelihood of becoming a member. They found that environmentally conscious young men with medium to low income, university-level education, living in the city center, and with lower levels of travel satisfaction are more likely to become a member of car-sharing services (Efthymiou and Antoniou, 2016)(Becker et al., 2017)(Prieto et al., 2017). Although focusing on the socio-demographic characteristics of those more prone to subscribe to car-sharing services in London, Madrid, Paris, and Tokyo, Prieto et al. (Prieto et al., 2017) found that local elements can impact individuals' likelihood of using car-sharing services. When examining the relevance of car-sharing services features, de Luca and Di Pace (de Luca and Di Pace, 2015) found that monetary travel cost, access time to car-sharing, and shared car availability are very important for switching to car-sharing in Italy. Yoon et al. (Yoon et al., 2017) found that monetary travel costs in relation to other transport alternatives are important for the decision of using car-sharing in Beijing (China), but access time and vehicle fuel type are not relevant. These findings highlight the importance of the local context when examining car-sharing subscription. As preferences for car-sharing services are likely to be affected by the local context, a car-sharing business model that can be perfect for one place are unlikely to be the best choice for all other places.

Aiming to understand how service-related features influence the decision to subscribe to car-sharing services and whether local context matters in these preferences, we collected and analyzed data from a stated-preference (SP) choice experiment conducted simultaneously in three cities: Copenhagen, Munich, and Tel Aviv-Yafo. By applying the same survey in the different cities, we are able to directly compare preferences across the three cities. The cities were chosen due to having diverse transport systems, norms, and mobility cultures. While Copenhagen has a strong bike culture (Københavns Kommune, 2020), Munich has an extensive and well-developed public transport network (Referat für Stadtplanung und Bauordnung - Landeshauptstadt München, 2017), and Tel Aviv-Yafo mobility relies on private cars and public buses (Sharav et al., 2018).

Finally, more recently, literature has shown the potential of offering incentives to nudge individuals' towards more sustainable mobility patterns and affect both the business success and the city at large (Azevedo et al., 2018). Whilst previous studies have focused on incentives for rebalancing the car-sharing fleet (Lippoldt et al., 2018) (Stokkink and Geroliminis, 2021), to date, no study has explored whether incentives can increase the likelihood of subscribing to car-sharing services. Identifying incentives that are potentially attractive for car-sharing users can help providers and urban stakeholders to increase service appeal in different contexts and to achieve multiple sustainability objectives. The contribution of this paper is twofold: (i) further examining the impact of different business models on car-sharing subscription in different contexts (cities); (ii) contribute to the literature of incentives for behavioral change by identifying relevant incentives to keep and attract car-sharing members.

## 2 Data and Methods
### 2.1 Case studies

The data used in this study was collected in Copenhagen, Munich, and Tel Aviv-Yafo. Copenhagen is the capital of Denmark and has a population of more than 1.2 million living in the Greater Copenhagen area. The first car-sharing scheme in Copenhagen was offered in 1998 at the request of the City of Copenhagen (Københavns Kommune). Subsequently, many car-sharing schemes have been established, typically as associations, and in 2014, the first free-floating car-sharing service was introduced. In 2020, a total of 192 parking spaces were reserved for station-based car-sharing cars in Copenhagen, being 7% of those destined for electric cars. It is worth mentioning that when buying a car in Denmark, a registration tax is required, which varies between 85%-150% of the taxable value of the car depending on the value of the car. Electric cars currently have a discount on the registration tax (Skat - Danish Customs and Tax Administration, 2020).

Munich is the capital of the state of Bavaria and Germany's third-largest city. It has approximately 2.6 million inhabitants, about half of them living within the city area and the other half living in suburban districts. In Munich, car-sharing is extremely popular. It was introduced in 1988, and in 2020 a total of 226 providers operate car-sharing at 840 locations in Germany (BCS - Bundesverband CarSharing, 2021). There is a motor vehicle tax obligation (Kraftfahrzeugsteuergesetz) in Germany for traffic on public roads.

Tel Aviv-Yafo has a population of over 450,000 people, is the second-largest city in Israel and the core of Israel's largest metropolis (~4,000,000 inhabitants). Since 2008, car-sharing services are offered in Tel Aviv-Yafo. Within the Tel Aviv-Yafo metro area, the service is currently provided in 5 cities of the inner ring – Tel Aviv-Yafo, Ramat Gan, Givatayim, Herzeliya, and Raanana; and is planned to expand to additional cities. According to Israel Central Bureau of Statistics (Central Bureau of Statistics, n.d.), in 2017, residents living in Tel Aviv-Yafo owned over 232,000 private cars, i.e., on average, half of the Tel Aviv-Yafo residents own a car. Taxation on most of the cars imported to Israel (no local manufacturing) reaches 83%. Hybrid cars are currently taxed at 30%, but this favorable taxation is being phased out. Gas in Israel is heavily taxed as well – about 65% of its value.

*2.2 Survey design*

The data used in this study was collected through a tailor-made online survey, which was designed based on the literature and the results from focus groups and interviews conducted in Copenhagen, Munich, and Tel Aviv-Yafo (Cantelmo et al. (Cantelmo et al., 2021)). The survey was implemented by combining a choice-based conjoint modeling tool Sawtooth (Sawtooth Software, 2021) and SPSS (IBM, n.d.) and made available online in both web and mobile versions. The survey was available in English and Danish for Copenhagen, in German and English for Munich, and in Hebrew and Arabic for Tel Aviv-Yafo. The general eligibility criteria were being 18 years or older and having a driver's license, except for Tel Aviv-Yafo, where the minimum age for using car-sharing services by the time of the survey was 21 years.  Questions to assess the eligibility of individuals were posed at the beginning of the survey, so respondents not eligible were screened out in the beginning. A small pilot was conducted, which led to improvements to the survey design, structure, and language.

The survey consisted of six parts. The first included a brief explanation and was followed by the second part, which had questions on socio-demographic information. In the third part, respondents were asked about their travel behavior and attitudes towards private cars and car-sharing services. The fourth consisted of questions to examine car-sharing incentives preferences where we provided a list of incentives but also gave the respondents the possibility of suggesting incentives not listed. The fifth part consisted of a Stated Preference (SP) experiment to understand the choice of respondents in regards to the subscription to different car-sharing plans. As attributes, we have included both incentives and attributes related to the characteristics of the different services as control variables (See Table 1).  Finally, as unfortunate events led to the survey being conducted during the outbreak of COVID-19, the sixth part consisted of questions to examine the effects of the COVID-19 pandemic on respondents' urban mobility behavior. For findings related to this part of the questionnaire, the reader is referred to Song et al. (Song et al., 2021). For the analysis in this paper, we use the data from parts 2, 4, and 5 of the questionnaire.

Table 1 SP Attributes and levels

| ATTRIBUTES | LEVELS | | | | | | | | | | | |
|---|---|---|---|---|---|---|---|---|---|---|---|---|
| | Copenhagen | | | | Munich | | | | Tel Aviv-Yafo | | | |
| | RT | OWST | OWFF | P2P | RT | OWST | OWFF | P2P | RT | OWST | OWFF | P2P |
| One-time subscription cost* | 200 kr | 200 kr | free | free | 50 € | 50 € | free | free | 50 ₪ | 50 ₪ | free | free |
| | 500 kr | 500 kr | 250 kr | 250kr | 100 € | 100 € | 50 € | 50 € | 130 ₪ | 130 ₪ | 65 ₪ | 65 ₪ |
| | 1000 kr | 1000 kr | 500 kr | 500 kr | 540 € | 540 € | 100 € | 100 € | 260 ₪ | 260 ₪ | 130 ₪ | 130 ₪ |
| Usage cost* | 1kr/min | 1kr/min | 1kr/min | 150kr/day | 0.19€/min | 0.19€/min | 0.19€/min | 20€/day | 0.25₪/min | 0.25₪/min | 0.25₪/min | 40₪/day |
| | 4kr/min | 4kr/min | 4kr/min | 200kr/day | 0.25€/min | 0.25€/min | 0.25€/min | 25€/day | 1₪/min | 1₪/min | 1₪/min | 52₪/day |
| | 6kr/min | 6kr/min | 6kr/min | 300kr/day | 0.39€/min | 0.39€/min | 0.39€/min | 30€/day | 1.6₪/min | 1.6₪/min | 1.6₪/min | 80₪/day |
| | 200kr/6h | 200kr/6h | 300kr/6h | 400kr/day | 1.5€/h | 1.5€/h | 35€/6h | 35€/day | 52₪/6h | 52₪/6h | 80₪/6h | 100₪/day |
| | 350kr/6h | 350kr/6h | 400kr/6h | 500kr/day | 2.5€/h | 2.5€/h | 13€/2h | 40€/day | 90₪/6h | 90₪/6h | 100₪/6h | 130₪/day |
| | 500kr/6h | 500kr/6h | 550kr/6h | 600kr/day | 6€/h | 6€/h | 18€/2h | 45€/day | 130₪/6h | 130₪/6h | 145₪/6h | 155₪/day |
| | 300kr/day | 300kr/day | 450kr/day | 800kr/day | 23€/day | 23€/day | 35€/day | 55€/day | 80₪/day | 80₪/day | 120₪/day | 210₪/day |
| | 500kr/day | 500kr/day | 650kr/day | 900kr/day | 35€/day | 35€/day | 49€/day | 70€/day | 130₪/day | 130₪/day | 170₪/day | 235₪/day |
| | 800kr/day | 800kr/day | 850kr/day | 1000kr/day | 48€/day | 48€/day | 79€/day | 80€/day | 210₪/day | 210₪/day | 220₪/day | 260₪/day |
| | up to 5 min | | | | | | | | | | | |
| | 6 to 10 min | | | | | | | | | | | |

| Walking time to access the vehicle | 11 to 15 min |
|---|---|
| Probability to get a shared vehicle | 10 out of 10 requests |
| | 9 out of 10 trip requests |
| | 7 out of 10 trip requests |
| Car-sharing vehicle types | One model of small city cars |
| | Small city cars and sedan cars |
| | Small, sedan and SUV cars |
| Car-sharing vehicle engine type | Combustion |
| | Electric |
| | Mix of combustion and electric |
| Walking time from the parking location to destination | up to 5 min |
| | 6 to 10 min |
| | 11 to 15 min |
| Extra features | Guaranteed child car seat availability |
| | Family/friends account with discounted rates |
| | A business account with discounted rates |
| | Booking in advance |
| | Plan including other modes for a seamless door to door trip |
| | Collect credits to redeem for goods (e.g., clothing and grocery discounts) |

RT: Round trip; OWST: One-way Station-based; OWFF: One-way free-floating; P2P: Peer-to-peer

\* Exchange rate (01st of September of 2020): 1 € = 7.4434 Kr. = 4.0183 ₪

The specification of attributes and levels of the SP experiment was made through an iterative process with representatives of all cities. The attributes included were one-time subscription cost, usage cost, walking time to access the vehicle, walking time from the parking location to the destination, the probability of getting a shared vehicle, car-sharing vehicle types, and car-sharing vehicle engine type. The incentives offered were guaranteed child car seat availability, a family/friends account with discounted rates, a business account with discounted rates, the possibility of booking in advance, a plan including other modes for a seamless door to door trip, and the possibility of collecting credits to redeem for goods (e.g., clothing and grocery discounts). We took into account existing services in each city, as well as their current features, prices, and packages, to define the levels. For the levels of the usage cost attribute, we explore different ways of presenting the cost (per minute, per hour, and per day) to verify whether this has an influence on respondents' choices. Each individual was presented with four tasks, the first three tasks designed on Ngene (Choice Metrics, 2010). The design was orthogonal and resulted in 108 scenarios, considering the attributes and levels defined above, that were grouped in 36 blocks of 3 tasks. As shown in Figure 1, each task presented four different car-sharing alternative plans (Roundtrip or RT, One-way Station-based or OWST, One-way Free-floating or OWFF, Peer-to-peer or P2P) and an opt-out alternative. The same design was presented to the respondents in each of the three cities, except for the costs that were defined and presented according to the local currencies and current service. The order of appearance of the attributes was random for each individual (but was the same across the three tasks of the same individual) in order to minimize response bias.

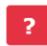

Figure 1 Example of choice task presented to respondents

## 2.3 Data collection

The data was collected from July 16th to August 06th of 2020 simultaneously in Copenhagen, Munich, and Tel Aviv-Yafo, and an additional data collection was performed in Munich between the 11th and 29th of September. At the time of the data collection, none of the cities was facing lockdown. For each city, a minimum sample size of 200 individuals was defined. The sampling strategy was opportunistic. While Copenhagen's and Munich's respondents were recruited through panels, in Tel Aviv-Yafo, respondents were contacted through different mailing lists[1].

---

[1] In Copenhagen, the panel comes from a survey company specialized in transport and, in Munich, the panel comes from the municipality of Munich and is composed of residents who previously stated to be willing to participate in studies about mobility. In Tel Aviv-Yafo, the recruitment was through posts on websites and newsletters from the municipality and car-sharing companies.

The completion rate in Copenhagen was 80%, while in Munich, it was 77%, and in Tel Aviv-Yafo, it was 39%. The relatively low completion rate in Tel Aviv-Yafo is believed to be associated with the different recruitment method used there and, consequently, the willingness that individuals contacted had on answering the survey.

*2.4 Sample characteristics*

After removing respondents that provided inconsistent answers and those who have answered the survey in fewer minutes than the 40% median, we ended up with a sample of 1276 valid respondents: 542 from Copenhagen, 490 from Munich, and 244 from Tel Aviv-Yafo. Inconsistent respondents were those who stated to be aware of car-sharing services in one question and later in the questionnaire answered that their lack of awareness about car-sharing is the reason why they do not use the service. Those who answered in less than 40% median time were removed from the sample because their short response times suggest a lack of attention and low data quality (Greszki et al., 2015). Table 1 presents the characteristics of the sample grouped by city.

More than 90% of the respondents in each of the cities are aware of car-sharing services. While the sample from Munich is the most balanced between current members and non-members, the sample from Copenhagen has more non-car-sharing members, and the sample from Tel Aviv-Yafo is mostly composed of car-sharing members due to the recruitment method. In all three cities, most of the respondents live in the main city, but Munich's sample has almost no respondents living in cities other than the main city (due to the recruitment method). Thus, our results for Munich reflect the preferences of those living in the city or suburbs and cannot be generalized for the metropolitan region; in Copenhagen, our findings reflect the preferences of non-members of car-sharing, and in Tel Aviv-Yafo, they reflect the preferences of car-sharing members.

Both Munich and Tel Aviv-Yafo samples have slightly more men than women, while the sample from Copenhagen is balanced in regards to gender (as targeted by the panel recruitment in Copenhagen). Munich and Tel Aviv-Yafo samples have more adults between 31 and 50 years old, whilst Copenhagen's sample has more respondents in the extreme categories (young and old ages), which is proportionally representative of the population (as targeted by the panel recruitment in Copenhagen).

As for the level of education, most respondents have at least a bachelor's degree, but the sample from Tel Aviv-Yafo is slightly more educated in general (88.83% has at least a bachelor's degree), and almost half the sample from Munich has at least a Master degree. The underrepresentation of the population that has up to a High school diploma or equivalent in Tel Aviv-Yafo is likely to be connected to the fact that car-sharing members in Tel Aviv-Yafo tend to be highly educated. In Munich, official statistics show high levels of education, suggesting that the distortion in the distribution of the education level in Munich's sample is limited. Most respondents from all cities have a high level of education, which is associated with higher rates of car-sharing usage (Dias et al., 2017) and higher acceptance of this mobility solution (Becker et al., 2017).

Additionally, most of the households have 1 or 2 members, and the majority of the respondents have up to one car, while more than a third of Tel Aviv-Yafo's sample respondents are from car-free households. This is likely to be a consequence of the fact that the majority of the respondents in Tel Aviv-Yafo are members of a car-sharing service (related to the recruitment method). Most of the respondents in all cities earn around the average or above, but Munich's sample has a lower number of respondents in the lower category, which is likely to be related to the overall high level of education of the individuals in the sample.

Table 1 Sample characteristics

|  | Copenhagen | | Munich | | Tel Aviv-Yafo | |
|---|---|---|---|---|---|---|
|  | Total | % | Total | % | Total | % |
| *Car-sharing membership status* | | | | | | |
| Car-sharing member | 95 | 17.53 | 225 | 45.92 | 156 | 63.93 |
| Past car-sharing member | 64 | 11.81 | 32 | 6.53 | 20 | 8.20 |
| Non-car-sharing member | 383 | 70.66 | 233 | 47.55 | 68 | 27.87 |
| *Gender* | | | | | | |
| Man | 266 | 49.08 | 284 | 57.96 | 134 | 54.92 |
| Woman | 275 | 50.74 | 203 | 41.43 | 108 | 44.26 |
| Prefer not to answer | 1 | 0.18 | 3 | 0.61 | 2 | 0.82 |
| *Age* | | | | | | |
| 18-30 | 145 | 26.75 | 58 | 11.84 | 36 | 14.75 |
| 31-40 | 88 | 16.24 | 158 | 32.24 | 88 | 36.07 |
| 41-50 | 97 | 17.90 | 147 | 30.00 | 63 | 25.82 |
| 51-60 | 88 | 16.24 | 71 | 14.49 | 36 | 14.75 |
| More than 60 | 124 | 22.88 | 56 | 11.43 | 21 | 8.61 |
| *Place of residence* | | | | | | |
| City center | 235 | 43.36 | 303 | 61.84 | 117 | 47.95 |
| Suburbs | 189 | 34.87 | 185 | 37.76 | 84 | 34.43 |
| Another city in the metropolitan region | 71 | 13.10 | 2 | 0.41 | 16 | 6.56 |
| Outside the metropolitan region | 47 | 8.67 | 0 | 0.00 | 27 | 11.07 |
| *Level of education* | | | | | | |
| Less Than High School | 39 | 7.20 | 22 | 4.49 | 2 | 0.82 |
| High school diploma or equivalent | 150 | 27.68 | 96 | 19.59 | 12 | 4.92 |
| Bachelor's degree | 169 | 31.18 | 52 | 10.61 | 97 | 39.75 |
| Master's degree | 134 | 24.72 | 181 | 36.94 | 77 | 31.56 |
| Doctoral degree | 8 | 1.48 | 57 | 11.63 | 12 | 4.92 |
| Other | 17 | 3.14 | 56 | 11.43 | 10 | 4.10 |
| Did not answer | 25 | 4.61 | 26 | 5.31 | 34 | 13.93 |
| *Number of cars in the household* | | | | | | |
| 0 | 139 | 25.65 | 162 | 33.06 | 112 | 45.90 |
| 1 | 303 | 55.90 | 244 | 49.80 | 86 | 35.25 |
| 2 | 91 | 16.79 | 71 | 14.49 | 37 | 15.16 |
| >2 | 9 | 1.66 | 13 | 2.65 | 9 | 3.69 |
| *Income (before taxes and other deductions)\** | | | | | | |
| Low | 82 | 15.1 | 32 | 6.5 | 56 | 22.9 |
| Medium | 140 | 25.8 | 219 | 44.7 | 46 | 18.8 |
| High | 221 | 40.7 | 146 | 29.8 | 96 | 39.4 |
| Did not answer | 100 | 18.4 | 93 | 19.0 | 46 | 18.9 |

\* Exchange rate (01st of September of 2020): 1.1987 USD = 1 EUR = 7.4434 DKK = 4.0183 ILS. Low income: Copenhagen = Up to 250.000 kr./year; Munich = Up to €29,999/year; Tel Aviv-Yafo= Below 11,000 ₪/month; Medium income: Copenhagen = 251-500.000 kr./year; Munich = €30,000 - €94,999/year; Tel Aviv-Yafo= About 11,000 ₪/month; High income: Copenhagen = Over 500.000 kr./year; Munich = €95,000 or more/year; Tel Aviv-Yafo= Above 11,000 ₪/month.

*2.5 Descriptive statistics*

For those who have reported awareness of car-sharing services, we have asked what the three most important aspects related to car-sharing use are. We have grouped the responses of members, past members, and non-members to get a sense of whether there are significant differences in their perceptions. Figure 2 shows the differences in the relevance of the different service attributes among groups of users and cities.

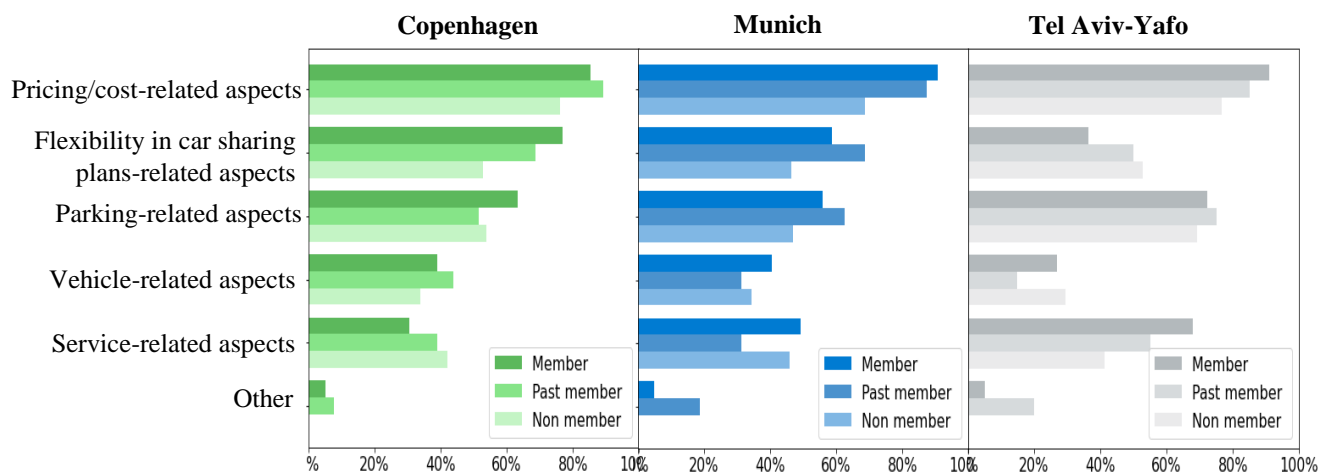

Figure 2 General preferences

Cost-related aspects are perceived as the most important for subscribing to a car-sharing service in all cities. While in Copenhagen and Munich, flexibility in the car-sharing plans is highly valued as well, in Tel Aviv-Yafo, more importance is given to service-related aspects, such as availability and coverage. Parking conditions are also highly relevant, especially for those from Tel Aviv-Yafo. Munich's car-sharing members also stated through "Other" that they believe it is important that the service shows measures towards climate protection, sustainability, and support to a social enterprise, and past members in Munich stated that it is important to have emission-free cars. Tel Aviv-Yafo's car-sharing members stated that car availability in the North Tel Aviv-Yafo area and the monthly membership cost are important, especially if it's an extra charge that is not deducted from the usage costs. As for usage patterns, in general, more than 30% of members and past members of car-sharing services reported they used the shared car for up to 30 min, and around 50% of them used the shared car for up to 1h.

## 2.6 Model specification

To examine individual's preferences for car-sharing plans and related incentives, we have estimated a joint mixed logit model with data from the three cities, accounting for correlation among choices of the same individual over the SP experiment (panel effect) (Train, 2003). As the variance of the error term (unobserved factors) is likely to vary among the three datasets (different cities)(Train, 2003), we have set the overall scale of utility by normalizing Copenhagen and included scale parameters ($\theta^c$) to allow for estimating the variances of Munich and Tel Aviv-Yafo relative to Copenhagen. By accounting for scale differences, we can compare the parameters from different datasets (Swait and Louviere, 1993). The utility specification is defined in Equation 1:

$$U_{int}^c = \theta^c(ASC_i^c + \beta_{ix}^c X_{int} + \beta_Z^c Z_n + \alpha_{in}^c + \sigma_{CSplans}^c + \varepsilon_{int}^c)$$
$$U_{0nt}^c = \theta^c(\varepsilon_{0nt}^c)$$
(1)

where $U_{int}^c$ is the utility that each individual n from city c associate to alternative i in the choice situation t and $ASC_i^c$ is the alternative specific constant, which captures the average effect on the utility of all factors not included in the model. $\beta_{ix}^c$ and $\beta_Z^c$ are the vectors of the coefficients associated with the impact of the service-related attributes included in the choice experiment ($X_{int}$) and the socioeconomic variables ($Z_n$) on the utility. Respectively $\alpha_{in}^c$ are error components normally distributed across individuals, which capture the correlation among choices for the same individual (panel effect), $\sigma_{CSplans}^c$ captures the magnitude of the correlation between the alternative car-sharing plans in each city and $\varepsilon_{int}^c$ is the i.i.d. extreme value error component. To perform the joint estimation of the models from each city, we defined fifteen alternatives (five for each city, namely: Roundtrip, One-way Station-based, One-way Free-floating, Peer-to-peer, and None of the alternatives). Each observation was associated with five alternatives, according to the city it comes from.

We have tested the interaction of the socio-demographic variables with both the alternative-specific constant and some of the attributes of the alternatives. The socio-demographic variables measured at the individual level were: gender, education (at least bachelor degree), occupation, age, car-sharing membership, car-sharing experience (members and past-members), and the socio-demographic variables measured at the household level were: children up to 12 years in the household, children up to 6 years in the household, car access, car ownership, number of cars, car leasing, number of cars, income and bike access. All socioeconomic variables tested were dummy variables, except for age that was entered in the model as a continuous variable, and the number of cars that was entered with the following levels: "0", "1 car", "2 or more cars". We tested alternative specific car-sharing membership variables to see whether being a member of the specific car-sharing service would impact the choice for that and alternative specific car access at home variables to examine whether those who have access to car perceived the alternatives differently. We also interacted the variables of incentives with the car-sharing membership variable to assess whether there was a difference in the preferences of members and non-members regarding the incentives proposed.

As for the socio-demographic variables, the final model includes only the coefficients of the variables or interactions that were found significant at least in one of the cities, namely age, children up to 12 years in the household, car-sharing membership, and income. Three income levels were included as dummy variables in the model, and the medium level was chosen as a reference. The definition of the income levels in each city can be found at the bottom of Table 1. We have also included a dummy variable missing income which was interacted with the constants to adjust the alternative specific constant of those individuals that did not answer the question on income.

All cost variables were converted to Euro (the exchange rate used can be found in the footnote of Table 1). We included binary variables to account for differences in choices due to the fact that prices were presented in different units across the alternatives and tasks (i.e., a charge per minute, per hour, or per day). Regarding car-sharing vehicle types available, the level one model of small city cars was used as the base level for vehicle type, while a mix of combustion and electric engine cars was the base level relative to car-sharing vehicle engine type. As the category small, sedan, and SUV car includes the category only small and sedan cars, we have tested interactions between them to try to isolate the effects of providing sedan in addition to small cars (reference) and the effects of providing SUV cars in addition to small and sedan cars. However, these interactions were not able to isolate impacts of the addition of each type of car (no significant coefficients) and, thus, were removed from the final model (the original variables without interactions were kept). As for the incentives offered, the reference level is the Business account with discounted rates, and the incentive Guaranteed child car seat availability was included in the model interacted with a dummy variable that took 1 if the respondent lives in a household with at least one kid less than 12 years old.

## 3. Results and Discussion

To estimate the model, we have tested the coefficients across the three cities for significant differences using both the likelihood ratio test and t-test. Where the model with restricted coefficients could not be rejected, and the coefficients were not significantly different at a 5% level, the coefficients of the cities were constrained to be the same (generic). One-time cost subscription was included as a generic coefficient across the three cities as the common preference parameter required for joint estimation. The joint model was estimated using PandasBiogeme (Bierlaire, 2020) and is presented in Table 2.

The alternative specific constants reveal a slight preference for peer-to-peer car-sharing in Copenhagen and Munich, while for one-way car-sharing in Tel Aviv-Yafo, everything else being equal. Significant panel effects indicate that the model captures the inherent correlations among the choices of the same respondent.

Table 2 Model results

| Variable | Copenhagen | | Munich | | Tel Aviv-Yafo | |
|---|---|---|---|---|---|---|
| | Estimate | Rob. Std err | Estimate | Rob. Std err | Estimate | Rob. Std err |
| $ASC$ - OWFF | 4.96*** | 1.06 | 6.37*** | 1.24 | 1.66*** | 0.604 |
| $ASC$ - OWST | 4.95*** | 1.06 | 5.8*** | 1.24 | 1.66*** | 0.593 |
| $ASC$ - P2P | 5.65*** | 1.07 | 6.53*** | 1.34 | 1.5** | 0.594 |
| $ASC$ - RT | 5.06*** | 1.06 | 5.23*** | 1.23 | 1.47** | 0.576 |
| $\alpha_{panel\ effect}$ - OWFF | 1*** | 0.189 | 1.97*** | 0.437 | -0.159 | 0.249 |
| $\alpha_{panel\ effect}$ - OWST | 1.04*** | 0.218 | 1.17* | 0.645 | 0.134 | 0.293 |
| $\alpha_{panel\ effect}$ - P2P | 0.818*** | 0.218 | 2*** | 0.489 | 0.612*** | 0.17 |
| $\alpha_{panel\ effect}$ - RT | 0.694*** | 0.25 | 2.91*** | 0.574 | 0.916*** | 0.204 |
| $\beta_{One\ time\ subscription\ cost}$ | -0.899*** | 0.101 | -0.899*** | 0.101 | -0.899*** | 0.101 |
| $\beta_{Usage\ cost\ (OWFF,\ OWST,\ RT)}$ | -0.092*** | 0.0246 | -0.619*** | 0.106 | -0.092*** | 0.0246 |
| $\beta_{Usage\ cost\ (P2P)}$ | -2.64*** | 0.33 | -10.1*** | 1.98 | -2.64*** | 0.33 |
| $\beta_{Usage\ cost\ per\ day}$ | -0.904*** | 0.148 | -0.904*** | 0.148 | -0.338*** | 0.0959 |
| $\beta_{Usage\ cost\ per\ hour}$ | -0.405*** | 0.146 | -0.213*** | 0.079 | -0.213*** | 0.079 |
| $\beta_{Only\ combustion\ cars}$ | -0.243*** | 0.0674 | -1.17*** | 0.258 | -0.243*** | 0.0674 |
| $\beta_{Only\ electric\ cars}$ | 0.0069 | 0.0989 | -0.123 | 0.202 | -0.0708 | 0.0614 |
| $\beta_{Only\ small\ and\ sedan\ cars}$ | -0.0282 | 0.108 | 0.54** | 0.225 | 0.0724 | 0.0715 |
| $\beta_{Small,\ sedan\ and\ SUV\ cars}$ | 0.151 | 0.105 | 0.205 | 0.212 | 0.117* | 0.0704 |
| $\beta_{Probability\ of\ finding\ a\ shared\ car}$ | 0.935** | 0.383 | 2.28*** | 0.745 | 0.374 | 0.228 |
| $\beta_{Walking\ time\ to\ access\ the\ vehicle}$ | -0.0101 | 0.0106 | -0.0317*** | 0.00907 | -0.0317*** | 0.00907 |
| $\beta_{Walking\ time\ from\ parking\ location\ to\ destination}$ | -0.0197*** | 0.00641 | -0.078*** | 0.0226 | -0.0197*** | 0.00641 |
| $\beta_{Incentive:\ Booking\ in\ advance}$ | 0.226* | 0.132 | 0.599** | 0.262 | 0.000352 | 0.0784 |
| $\beta_{Incentive:\ Guaranteed\ child\ car\ seat\ availability}$ | 0.469* | 0.276 | 1.35*** | 0.517 | 0.108 | 0.138 |
| $\beta_{Incentive:\ Collect\ credits\ to\ redeem\ for\ goods\ (e.g.,\ clothing\ and\ grocery\ discounts)}$ | 0.16 | 0.131 | 0.0581 | 0.27 | -0.0555 | 0.0851 |
| $\beta_{Incentive:\ Family/friends\ account\ with\ discounted\ rates}$ | 0.19 | 0.134 | 0.75*** | 0.271 | 0.0124 | 0.0867 |
| μ<sub>Incentive: Plan including other modes for a seamless door to door trip</sub> | 0.3** | 0.134 | 0.453* | 0.268 | 0.0233 | 0.0917 |
| σ<sub>Incentive: Plan including other modes for a seamless door to door trip</sub> | | | | | 0.366* | 0.21 |
| β<sub>Age</sub> | -1.04*** | 0.163 | -1.04*** | 0.163 | -0.122 | 0.0744 |
| β<sub>Car-sharing membership</sub> | 1.15* | 0.682 | 2.2*** | 0.792 | 0.857*** | 0.27 |
| β<sub>High income – household</sub> | -0.48 | 0.64 | -1.17 | 0.842 | -0.608** | 0.304 |
| β<sub>Low income – household</sub> | 0.472 | 0.845 | 0.878 | 1.55 | -0.816** | 0.357 |
| β<sub>Missing income - household</sub> | -0.218 | 0.813 | -2.33** | 1.01 | -0.858** | 0.36 |
| β<sub>Household with children up to 12 years</sub> | 1.66*** | 0.556 | 1.66*** | 0.556 | -0.212 | 0.2 |
| σ<sub>CSplans</sub> | 4.81*** | 0.443 | 5.8*** | 0.919 | 0.935*** | 0.223 |
| Scale (Θ) [a] | | | 0.543*** | 0.0746 | 2.31*** | 0.45 |
| Number of observations | 3737 | | | | | |
| Number of individuals | 1276 | | | | | |
| Number of draws | 5000 | | | | | |
| Number of estimated parameters | 88 | | | | | |
| Log-likelihood | -4796.552 | | | | | |
| Null log-likelihood | -6014.469 | | | | | |
| Rho-square | 0.202 | | | | | |
| Adjusted rho-square | 0.188 | | | | | |

\*\*\* Significant at 1% level
\*\* Significant at 5% level
\* Significant at 10% level
[a] T-test against 1

The cost coefficients indicate that this attribute negatively affects the likelihood of subscribing to a car-sharing plan, which is consistent with the behavioral theory. An increase in the usage cost is perceived more negatively for those living in Munich than for those living in Copenhagen and Tel Aviv-Yafo. This may be a consequence of the fact that the large majority of individuals in the German sample lives in Munich, whilst the Danish and Israeli samples include more individuals who live in other cities of the metropolitan region and even outside it, who are likely to have lower public transport access levels and general higher transportation costs. Moreover, respondents from Copenhagen and Tel Aviv-Yafo face higher taxation when buying a private car.

When accounting for the presentation of costs for the different car-sharing plans as packages paid per minute, per hour, and per day, results indicate that the payment per minute (reference level) is preferred, followed by hourly thereafter daily rates. Individuals' intended car-sharing usage patterns may have an effect on that, i.e., they expect to use it for relatively short trips. It should be noted that more than 30% of members and past members of car-sharing services reported they used the shared car for up to 30 min and around 50% of them used the shared car for up to 1h.

To account for accessibility at origin and destination, we have included attributes related to both walking time to access the vehicle and walking time from the parking location to the destination. For Munich and Tel Aviv-Yafo respondents, the higher the walking time to access the shared car, the lower the probability of subscribing to a car-sharing plan. Overall, our results suggest that Munich's respondents are more sensitive to walking times and respondents from Copenhagen's are the least sensitive to this aspect of the service, which may be related to a stronger active mode culture. The attachment of high importance to the costs and access times was also found in de Luca and Di Pace (de Luca and Di Pace, 2015) study.

As for preferences related to shared cars fuel type, the results indicate that there is no significant difference in individuals' preference for a service with a fleet composed of a mix of combustion and electric engine cars (reference level) and a fleet composed only by electric vehicles. This suggests high public acceptance for a shared electric cars only fleet. However, especially in Munich, having a fleet composed of only combustion cars negatively affects the probability of subscribing to a car-sharing plan, which is an indication of stronger environmental concern. In regards to vehicle type, respondents from Munich prefer services with fleets composed of not only small city cars (reference level) but also sedan cars, while those from Tel Aviv-Yafo prefer varied fleets with small, sedan, and SUV cars. As expected, the probability of finding a shared car, which is related to their availability, positively affects the likelihood of respondents from Copenhagen, Munich, and Tel Aviv-Yafo to subscribe to a car-sharing plan.

In regards to preferences for car-sharing incentives, these are measured relative to the incentive business account with discounted rates, which is the reference level. In Copenhagen and Munich, the incentive booking in advance is significantly preferred over the reference, while in Tel Aviv-Yafo, the preference for booking in advance is not significantly different from the reference. The same is true for the incentive guaranteed child car seat availability, which is interacted with the socio-demographic variable household with children up to 12 years. The preference for the incentive related to credits to redeem for goods was not found to be significantly different from the reference in all three cities. The incentive family/friends account with discounted rates is not significantly different from the incentive business account (reference) for respondents from Copenhagen and Tel Aviv-Yafo but is preferred over the latter for those living in Munich. Finally, the incentive plan including other modes for a seamless door-to-door trip is preferred over the reference for respondents in Copenhagen, Munich, and most respondents from Tel Aviv-Yafo.

Concerning the influence of socio-demographic characteristics in the likelihood of subscribing to a car-sharing plan, the age parameters in Copenhagen and Munich are negative, suggesting that the older an individual is, the less likely to subscribe to a car-sharing plan. This finding is in line with Prieto et al. (Prieto et al., 2017), who argue that possible explanations are long-term private car use habit and generation effects. The variables related to income were only significant for Tel Aviv-Yafo, indicating that individuals with high and low income have a lower probability of subscribing compared to those with medium income (reference level). This finding in Tel Aviv-Yafo is aligned with Efthymiou and Antoniou (Efthymiou and Antoniou, 2016); however, there is no clear agreement on the literature about whether high-income increases (Giesel and Nobis, 2016)(Yoon et al., 2017)(Dias et al., 2017) or decreases the likelihood of subscribing and using car-sharing (Efthymiou and Antoniou, 2016)(Zhou and Kockelman, 2011). It seems that other contextual variables may play a role in how individuals with different economic profiles perceive car-sharing services as, for example, residential location, which is highly correlated with income in some cities. More agreement exists on the tendency of car-sharing subscription and usage among low-income individuals, as they have been associated with lower chances of using car-sharing services, perceiving it as expensive (Efthymiou and Antoniou, 2016)(Dias et al., 2017). Both in Copenhagen and Munich, individuals living in households that have a child up to 12 years are more likely to subscribe to a car-sharing plan. Although most literature supports that individuals living in a household with children were less likely to subscribe and use car-sharing services (Efthymiou and Antoniou, 2016)(Dias et al., 2017), the mobility biographies literature supports that when individuals face life course changes (specifically childbirth and/or children starting at daycare, kindergarten, or school) they are more likely to start using car-sharing (Priya Uteng et al., 2019). Not surprisingly, those who are already car-sharing members were more likely to choose one of the plans offered, as opposed to the opt-out alternative (not subscribing to a car-sharing plan). This was expected because car-sharing membership indicates a predisposition and underlying preference for the service, as well as experience with the service and its attributes.

Finally, the significant error components show that the model is capturing the correlation between the alternative plans in each city, and the scales indicate that the variance of unobserved factors is greater in Munich than in Copenhagen and lower in Tel Aviv-Yafo than in

Copenhagen.

## 4 Conclusions

We examined individuals' preferences towards different features and incentives associated with car-sharing services in Copenhagen, Munich, and Tel Aviv-Yafo. The model reveals that the local context indeed affects individuals' perceptions and preferences. Although some car-sharing features are likely to be relevant everywhere (e.g., pricing, availability of shared cars), the local context has an effect on the preferences for other features. In general, offering good-priced packages and good availability of shared cars are essential for members and potential members.

In Munich, the results indicate that car-sharing fleet composition in terms of both vehicle type and fuel type is highly relevant. While the former suggests that members and potential members anticipate the need for extra trunk space (related to intended purposes for using the shared cars), the latter is likely to be a consequence of the relatively higher environment concern those living in Munich display. Marketing campaigns focusing on the positive environmental consequences of car-sharing usage are expected to be highly appealing there, especially if supported by data and studies on the actual impacts it has in Munich. Moreover, the accessibility of shared cars (walking times) is highly appreciated in Munich's market, revealing that municipalities and operators need to work closely to strategically coordinate the provision of parking spaces and recharging infrastructure together with the car-sharing network.

Beyond good alternative packages and availability of shared cars, providing good parking conditions near destinations is an important aspect in the Copenhagen car-sharing market. This suggests that while individuals do not see it as a huge effort having to walk up to around 15 minutes to access an available shared car, being able to park within a reasonable distance from the destination is essential. This indicates that municipalities should give special consideration to the popularity of the location when deciding on where to allocate reserved parking spaces.

For the Tel Aviv-Yafo market, providing good parking conditions and a varied shared cars fleet with small, sedan, and SUV cars are highly appreciated. The analysis of the data and results suggests a relatively higher parking pressure in Tel Aviv-Yafo and that an adequate provision of reserved parking spaces for shared cars is essential for the increasing service ridership (both by the municipalities and by relevant private commercial locations). The preference for a more varied car-sharing fleet indicates that respondents (mainly members) use shared cars for different purposes and that marketing campaigns presenting the different uses for shared cars may attract more subscribers.

The analysis of the relevant socio-demographic variables reveals the potential of marketing campaigns targeting young individuals with children in Copenhagen and Munich and mid-income individuals in Tel Aviv-Yafo. As discussed, pricing can be a barrier for low-income individuals in Tel Aviv-Yafo, suggesting the need to explore different pricing schemes.

In regards to individuals' preferences for incentives, in general, both in Copenhagen and Munich, individuals highly value booking in advance, guaranteed child car seat availability, and plan including other modes for a seamless door-to-door trip. Additionally, the possibility of having a plan including other modes for a seamless door-to-door trip is preferred by most respondents in Tel Aviv-Yafo, and the opportunity of a family/friends account with discounted rate is also highly relevant in Munich's car-sharing market.

The results improve our understanding of how both the features of the car-sharing business model and the provision of incentives can maintain and attract members to the system. The achieved insights pave the road for the actual design of car-sharing business models and incentives that can be offered by existing and future car-sharing companies in the studied or similar cities. Our findings also indicate the market segments with a higher likelihood to join car-sharing services in each city, which can be explored by local car-sharing operators.

This paper has some limitations. It should be noted that the sample from Germany is mainly composed of respondents living in Munich, and the results do not reflect the mobility needs of those living outside the city. Their preference for shorter walking times, particularly, may be a consequence of better public transport accessibility in the main city. Moreover, the sample from Tel Aviv-Yafo and Copenhagen are not balanced in terms of car-sharing membership. Further research is needed to check the stability of preferences across different areas and groups with different membership statuses. Additionally, as most respondents from all cities are highly educated, they are likely to display higher acceptance of car-sharing plans (as opposed to having chosen the "None of the alternatives" option) than other segments of the population. A natural extension of this research, which can potentially account for this limitation, is the inclusion of attitudinal variables in the model to control for car-sharing attitudes. Such extension is under development as part of our ongoing research. Finally, we collected data in three cities, which enriched our understanding of contextual differences, but the replication of the study in other cities (and continents) has the potential to expand our perspective on the differences and similarities of the car-sharing markets across the world.

**Acknowledgements**

This work was supported by the EIT Urban Mobility through the research project Shared Mobility Rewards (ID: 20045). We would like to thank Epinion, Car2Go, AutoTel, and the

<em>Presented at the Transportation Research Board Annual Meeting - January 9–13, 2022; Washington, DC</em>